\documentclass[11pt,a4paper]{article}
\pdfoutput=1
\usepackage{jheppub}

\usepackage{braket, bm}

\newcommand{\cl}{\mathrm{cl}}
\newcommand{\q}{\mathrm{q}}
\newcommand{\ud}{\mathrm{d}}
\newcommand{\AI}{\mathcal{I}}

\title{Statistics on Lefschetz thimbles:~Bell/Leggett-Garg inequalities and the classical-statistical approximation}

\author[a]{Peter Millington,}
\author[c,b]{Zong-Gang Mou,}
\author[a]{Paul M. Saffin,}
\author[b]{Anders Tranberg,}

\affiliation[a]{School of Physics and Astronomy, University Park, University of Nottingham,\\ Nottingham NG7 2RD, United Kingdom}
\affiliation[b]{Faculty of Science and Technology, University of Stavanger, 4036 Stavanger, Norway}
\affiliation[c]{Department of Physics and Astronomy, Rice University, Houston, Texas 77005, U.S.A.}

\emailAdd{p.millington@nottingham.ac.uk}
\emailAdd{zm17@rice.edu}
\emailAdd{paul.saffin@nottingham.ac.uk}
\emailAdd{anders.tranberg@uis.no}

\abstract{
Inspired by Lefschetz thimble theory, we treat Quantum Field Theory as a statistical theory with a complex Probability Distribution Function (PDF). Such complex-valued PDFs permit the violation of Bell-type inequalities, which cannot be violated by a real-valued, non-negative PDF. In this paper, we consider the Classical-Statistical approximation in the context of Bell-type inequalities, viz.~the familiar (spatial) Bell inequalities and the temporal Leggett-Garg inequalities. We show that the Classical-Statistical approximation does not violate temporal Bell-type inequalities, even though it is in some sense exact for a free theory, whereas the full quantum theory does. We explain the origin of this discrepancy, and point out the key difference between the spatial and temporal Bell-type inequalities.  We comment on the import of this work for applications of the Classical-Statistical approximation.
}

\begin{document}

\maketitle

%%%%%%%%%%%%%%%%%%%%%%%%%%%%%%%%%%%%%%%%%%%%%%%%%%%%%
\section{Introduction}
\label{sec:Intro}

\noindent What is the obstacle that generally hinders computations in Quantum Field Theory (QFT), when we can readily perform the corresponding computations in Quantum Mechanics (QM)?
The short answer is the infinite number of degrees of freedom in QFT.
To see how the degrees of freedom may be enumerated, we can look at the functional Schr\"odinger equation for the real, interacting Klein-Gordon field $\phi$. This takes the form
\begin{align}
\label{eq:schr}
i\hbar \frac{\partial}{\partial t} \langle \phi,t | {\rm in} \rangle = \int \ud^d\mathbf{x} \left[-\frac{\hbar^2}{2}\frac{\delta^2}{\delta \phi^2(t,\mathbf{x})} + \frac{1}{2}\left(\bm{\nabla} \phi(t,\mathbf{x})\right)^2+V(\phi(t,\mathbf{x}))\right] \langle \phi,t | {\rm in} \rangle~,
\end{align}
where $V$ is the potential and $\langle \phi,t | {\rm in}\rangle $ is the field-configuration representation of the wavefunctional with initial state $|{\rm in}\rangle$. Each $\langle \phi,t | {\rm in} \rangle$ corresponds to a cross-section of $\phi$ line bundle, and the total number of cross-sections would amount to $N_{\phi}^{(N_{x})^d}$, if we discretized both the $\mathbf{x}$ and $\phi$ spaces. Notice the appearance of the number of spatial dimensions $d$ in the exponent, which indicates an exponential growth in the number of degrees of freedom with $N_{x}$. The special case, of course, is $d=0$, which is, in fact, QM.
Furthermore, the evaluation of the functional derivative, $\delta^2/\delta \phi^2(t,\mathbf{x})$, makes use of the neighbouring states/cross-sections. Thus, to solve the functional Schr\"{o}dinger equation, we should keep the values for all of them, and the exponential growth in the number of degrees of freedom renders the approach insoluble on a normal computer.

While, in quantum theory, all states are equal, some states are more equal than others.
If we do the statistics correctly, we do not really need to include all states in order to get an accurate enough result. This is the idea behind the stochastic method.  The question is: What determines whether one state is more important than another, or, in other words, from which Probability Distribution Function (PDF) should we sample these states? The answer is well known in QFT and it can be extracted from a path-integral approach to the problem.

Recall that, to calculate the expectation value \smash{$\braket{\hat{\mathcal{O}}(t)}$} of some operator \smash{$\hat{\mathcal{O}}(t)$}, we compute \smash{$\braket{\hat{\mathcal O}} = {\rm Tr}\big[\hat{\rho}(t_0)\hat{\mathcal{O}}(t)\big]/{\rm Tr}\big[\hat{\rho}(t_0)\big]$},  where we assume the density operator \smash{$\hat{\rho}$} is known at the time $t_0$.
The partition function ${\rm Tr}\left[\hat{\rho}(t_0)\right]$ is the integral of the PDF. For an analogy, consider the normalization \smash{$\int \ud x \;e^{-x^2}$}, arising when calculating \smash{$\braket{x^2} = \int \ud x \;e^{-x^2} x^2/\int \ud x \;e^{-x^2}$} for the Gaussian PDF \smash{$e^{-x^2}$}.
In the expression for $\langle \hat{\mathcal O}(t)\rangle$, the initial density operator $\hat{\rho}(t_0)$ and the operator $\hat{\cal O}(t)$ are evaluated at different times.  Therefore, to construct a path-integral representation of this expectation value, we should connect these operators together by constructing a closed-time path integral, according to the Schwinger-Keldysh prescription \cite{Schwinger:1960qe,Keldysh:1964ud}. More specifically, we can write the expectation value as
\begin{align}\label{eq:expectation_O}
&\langle\hat{\cal O}(t)\rangle=\frac{
\int {\mathcal D} \phi  
\langle\phi^+_0;t_0|\hat\rho(t_0)|\phi^-_0;t_0\rangle e^{ \frac{i}{\hbar}\int_{\cal C}\ud t'\,L } {\mathcal O}(t)}
{\int {\mathcal D} \phi  
\langle\phi^+_0;t_0|\hat\rho(t_0)|\phi^-_0;t_0\rangle e^{ \frac{i}{\hbar}\int_{\cal C}\ud t'\,L }}~,
\end{align}
where $\mathcal{D}\phi\equiv \mathcal{D}\phi^+\,\mathcal{D}\phi^-$ is the functional integration measure, $L$ is the Lagrangian and
$\mathcal{O}(t)$ is the field-configuration representation of the operator.
The ${\mathcal C}$ on the time integrals indicates the closed-time contour: a contour with a time-ordered ``$+$'' branch, starting from $t_0$ and ending at some time $t_m>t>t_0$, and an anti-time-ordered ``$-$'' branch running backwards from $t_m$ to $t_0$.  The superscripts ``$+$'' and ``$-$'' on the initial field configuration $\phi_0$ identify whether its time argument lies on the $+$ or $-$ branch of the closed-time contour $\mathcal{C}$.

Following the Schwinger-Keldysh approach, the expression~\eqref{eq:expectation_O} can be obtained directly by 
using the time-evolution operator \smash{$\hat{U}(t_1,t_2)=\mathrm{T}\exp\big[-\frac{i}{\hbar}\int_{t_1}^{t_2}\ud t'\,\hat{H}(t')\big]$}, where $\hat{H}$ is the Hamiltonian operator and $\mathrm{T}$ is the time-ordering operator,\footnote{In the absence of external sources, the Hamiltonian operator is time-independent in the Heisenberg picture.  Here, we have allowed for explicit time dependence, e.g., through the parameters of the theory.}
or by inserting complete sets of eigenstates of the Heisenberg-picture field operators.
We note that the choice of the end time of the contour $t_m$ is arbitrary, so long as $t_m>t$. Alternatively, one can arrive at the same expression from the functional Schr\"{o}dinger equation \eqref{eq:schr} by recalling that its solution can be written as a convolution of the path integral with the initial wavefunctional, i.e.,~\smash{$\Psi(\phi_m;t_m)=\int^{\phi_m}{\mathcal D}\phi\, e^{iS/\hbar}\Psi(\phi_0;t_0)$}, with the expectation value computed as \smash{$\braket{\hat{\mathcal{O}}(t)} = \int {\mathcal D}\phi\, \Psi^*(\phi_t^-;t)\mathcal{O}(t) \Psi(\phi_t^+;t)$}.

With this path integral in mind, our plan to compute expectation values by picking a sample of important states is analogous to doing an integral using generated samples, such as one might do in a Monte-Carlo approach.
However, by rephrasing QFT in the language of statistics, we are not led to any direct solution.  In fact, we face a more challenging question than is found in standard statistics: How to deal with the fact that the PDF is complex-valued.

The non-negativity of the PDF in standard statistical analysis has a number of crucial effects.  On one hand, non-negativity is an important condition to guarantee a well-behaved Markov chain, thereby allowing a Monte Carlo computation.
On the other hand, there are some key properties that are implied by non-negativity.
One of them relates to the Bell inequalities~\cite{Bell:1964kc}. For instance, from the viewpoint of statistics, there can be no violation of the inequality
\begin{align}
\label{eq:genineq}
-3\leq \langle A B\rangle + \langle B D\rangle - \langle A D\rangle \leq 1~,
\end{align}
if the observables $A$, $B$ and $D$, of value $\pm 1$, are drawn from some real-valued, non-negative distribution function. One can prove this by generating samples according to such a distribution.  Since any single realisation will satisfy the inequality, so will the average, given that $\braket{A B} =\sum_{i=1}^N A_iB_i/N$ and $A_i$, $B_i$ and $D_i$ are restricted to be $\pm 1$. However, a complex PDF can lead to violations of such inequalities.

Complex PDFs call for complex analysis. Consider, e.g., the complex Gaussian integral \smash{$\int\ud x\,e^{ix^2}$}. With the complex PDF $e^{ix^2}$, we cannot ascertain which states are more probable than others, unlike with the normal distribution $e^{-x^2}$. Nevertheless, we know that we can perform the integral by deforming the contour of integration into the complex plane by Cauchy's theorem.  By choosing a better contour, we can then make sense of the statistics. In this simple case, we find a normal distribution if we rotate the integration contour by $\pi/4$ in the complex $x$ plane, so that $ix^2\to-x^2$. Under the same principle, the Lefschetz thimble approach provides a powerful tool for QFT, whereby we complexify all real-valued fields \cite{Witten:2010cx} and are furnished with a prescription for finding a suitable integration contour.
The application of the Lefschetz thimble approach to real-time path integrals has recently attracted significant attention~\cite{Tanizaki:2014xba, Cherman:2014sba,Alexandru:2016gsd, Alexandru:2017lqr, Mou:2019tck, Ai:2019fri, Mou:2019gyl, Alexandru:2020wrj}.  

A popular approximation used in the evaluation of the real-time path integral is the Classical-Statistical (CS) approximation. There, the quantum evolution is replaced by the classical evolution of an ensemble of initial conditions, drawn from a non-negative PDF. The interpretation in the context of the Lefschetz thimble Monte Carlo is that we keep only the critical points of the action and allow only for non-negative Wigner functions.  This is quantitatively a good approximation in the limit where field occupation numbers are large. Its validity may be further expressed in terms of discarding diagrams in a perturbative expansion (at high temperature \cite{Aarts:1997kp}), in terms of the ``statistical'' propagator being much larger than the ``spectral'' propagator (in direct comparison with quantum evolution based on truncations of the 2PI effective action \cite{Aarts:2001yn,Arrizabalaga:2004iw,Rajantie:2006gy,Berges:2007ym,Berges:2010nk}) or in terms of the state being ``squeezed'' (e.g., in cosmological perturbations during inflation \cite{Mukhanov:1990me}).

The prescription is as follows: When field modes acquire large occupation numbers, one may evolve them using classical equations of motion and compute observables as in classical field theory. When the fields interact, all modes are coupled, but the dynamics is still expected to be dominated by the ones with high occupancy. In a few specific cases, one may opt to seed the evolution with an initial state resembling the quantum zero-point fluctuations (the ``half'' \cite{Rajantie:2000nj,GarciaBellido:2002aj,Smit:2002yg}, and for critical analyses, see Refs.~\cite{Braden:2018tky, Hertzberg:2020tqa}). This assumes that the evolution is linear (non-interacting) until the phenomenon under consideration (e.g., resonance, instability or inflation) amplifies these seeds into large occupation numbers, $\gg 1$. The ``half'' can of course not lead to truly quantum phenomena and is, from the point of view of the classical evolution, simply a curious non-thermal initial condition.
Even so, it can give rise to spurious effects, since it has divergent energy in the UV (see, for instance, Ref.~\cite{Arrizabalaga:2004iw}).

In the case of a free theory, the CS approximation is, in some sense, exact.  In spite of this, the aim of this article is to describe where and why the CS approximation is nevertheless incapable of describing the violation of Bell-type inequalities, both spatial Bell inequalities and temporal Leggett-Garg inequalities, even for a free theory. 

The remainder of this article is organised as follows.  In Sec.~\ref{sec:CSinLefschetz}, we provide a brief introduction to the CS approximation within the context of the Lefschetz thimble approach.  In Secs.~\ref{sec:LG} and~\ref{sec:Bell}, we focus on the temporal Leggett-Garg inequalities and the spatial Bell inequalities, respectively.  Our conclusions are presented in Sec.~\ref{sec:conc}.  Some useful results are collected in the Appendices.

%%%%%%%%%%%%%%%%%%%%%%%%%%%%%%%%%%%%%%%%%%%%%%%%%%%%%

\section{Classical-Statistical approximation and Lefschetz thimbles}
\label{sec:CSinLefschetz}

To compute the expectation value appearing in Eq.~\eqref{eq:expectation_O}, it proves convenient to introduce the following field variables:\footnote{We adopt the coefficients used in Refs.~\cite{Greiner:1996dx, Aarts:1997kp} but the notation used in Ref.~\cite{Kamenev:2009}. For clarity, we suppress the spacetime index for the fields, except in the places where the specific spacetime sites are required.}
\begin{align}
\label{eq:phicl_phiq}
\phi^{\cl}\equiv\frac{1}{2}\left( \phi^++\phi^- \right)~,
\qquad\phi^{\q}\equiv\phi^+-\phi^-~.
\end{align}
In terms of these variables, the closed-time path integral involves the exponent
\begin{align}
\frac{i}{\hbar}\int_{\cal C}\ud t\;L
&=\frac{i}{\hbar} \int_{\partial {\mathcal C}} \ud^d \mathbf{x} \left[-\phi_0^{\q}\dot{\phi}_0^{\cl} \right]-\mathcal{I}~,
\label{eq:L}
\end{align}
with the bulk term
\begin{equation}
\label{eq:I}
\mathcal{I}=
\frac{i}{\hbar} \int \ud^{d+1} x \left[ \phi^{\q}\left(\ddot{\phi}^{\cl} - \bm{\nabla}^2 \phi^{\cl} +V^{(1)} \right)
+\sum_{n=1}^{+\infty} \frac{\left(\phi^{\q}\right)^{2n+1}}{ 2^{2n}(2n+1)!} V^{(2n+1)} 
\right]~,
\end{equation}
where $V^{(k)}\equiv \ud^k V(\phi)/\ud \phi^k|_{\phi\to\phi^{\cl}}$.  The partition function can then be arranged in the form
\begin{align}
Z= \int {\mathcal D} \phi_0^{\cl}\, {\mathcal D} \pi_0^{\cl}\,W\big(\phi_0^{\cl},\pi_0^{\cl};t_0\big)\int \mathcal{D}'\phi^{\q}\, \mathcal{D}'\phi^{\cl}\; e^{-\AI}~,
\label{eq:Z}
\end{align}
where
\begin{align}
\label{eq:Weyl}
W\big(\phi^{\cl}_0,\pi^{\cl}_0;t_0\big) \equiv \int {\mathcal D}\phi^{\q}_0\;\Big\langle \phi^{\cl}_0+\tfrac{\phi^{\q}_0}{2} \Big | \hat\rho(t_0)  \Big| \phi^{\cl}_0-\tfrac{\phi^{\q}_0}{2} \Big\rangle e^{-\frac{i}{\hbar}\int \ud^d \mathbf{x}\; \phi^{\q}_0 \pi^{\cl}_0}~,
\end{align}
is the initial Wigner function, in which the boundary term from Eq.~\eqref{eq:L} has fulfilled the role of the kernel in a Weyl transform of the initial density operator.  Notice that we have relabelled $\pi_0^{\cl} \equiv  \dot\phi_0^{\cl}$ to make contact with the Hamiltonian form of the path integral (see Ref.~\cite{Hertzberg:2019wgx}). The prime on the integration measure indicates that the fields on the initial temporal boundary, such as $\phi_0(\mathbf{x})$ and $\pi_0(\mathbf{x})$, have been excluded.

Due to the Hermiticity of the density operator, i.e., $\hat\rho^\dagger=\hat\rho$, the initial Wigner function must be real-valued.
In comparison, the bulk term $e^{-\AI}$ is purely a phase term.
Notice in addition that, to make the bulk path integral well-defined, the existence of $\ddot \phi^{\cl}$ requires two temporal boundary conditions, which are provided here by the Wigner function through $\phi_0^{\cl}$ and $\dot\phi_0^{\cl}$.
This structure also has an impact on the critical point and related Lefschetz thimble as follows.

We first notice that the critical point\footnote{The critical point is actually a spacetime field configuration. We refer the reader to Refs.~\cite{Mou:2019tck, Mou:2019gyl} for the concrete derivation of the saddle point. The subtlety lies at the turning point.} of $\AI$ satisfies 
\begin{align}
\phi^{\q}=0 
\qquad \text{and} \qquad
 -\ddot{\phi}^{\cl} + {\bm \nabla}^2 \phi^{\cl} -V^{(1)}=0~,
\end{align}
with the initial values of $\phi^{\cl}$ and $\dot\phi^{\cl}$ determined by the initial Wigner function $W(\phi_0^{\cl},\pi_0^{\cl};t_0)$.
That is, the critical point corresponds to a $\phi^{\cl}$ that follows the classical trajectory with initial data specified by $W(\phi_0^{\cl},\pi_0^{\cl};t_0)$, and, as an initial value problem, there exists one and only one solution. This is the virtue of the two-step evaluation of the path integral. Namely, if we separate the path integral into the initial Wigner function $W(\phi_0^{\cl},\pi_0^{\cl};t_0)$ and the dynamical part $e^{-\AI}$, there will exist one and only one Lefschetz thimble for each initialization generated by the Wigner function. By Lefschetz thimble, we mean the manifold generated by the gradient flow originating from the critical point \cite{Witten:2010cx}, and therefore the number of thimbles equals the number of critical points.
For more on the two-step evaluation, we refer the reader to Refs.~\cite{Mou:2019tck,Mou:2019gyl}.

The dynamical part $\AI$ consists of odd terms in $\phi^{\q}$.
If there are only linear terms in $\phi^{\q}$, we can integrate $\phi^{\q}$ out to obtain functional delta functions in $\phi^{\cl}$. Conversely, for nonlinear potentials that will yield odd terms in $\phi^{\q}$ of higher powers, imposing the same delta functions corresponds to the approximation of dropping these higher terms from Eq.~\eqref{eq:I}. This is known as the Classical-Statistical (CS) approximation~\cite{Moore:1996bn, Aarts:1997kp, Cooper:2001bd, Berges:2006xc, Epelbaum:2014yja}. Note that this approximation maintains the non-linearity in $\phi^{\cl}$.
In corollary, no such approximation is needed if there are only quadratic terms in the Lagrangian, and the CS approximation is then, in some sense, exact. The situation remains, however, non-trivial when the parameters vary in space and time.  
For instance, certain quantum effects in the early Universe can be well approximated via quadratic terms on the time varying background, and the ensemble average of classical evolutions will provide an honest description.

After integrating out $\phi^{\q}$, the CS approximation leads to 
\begin{align}
Z= \int {\mathcal D} \phi_0^{\cl}\, \pi_0^{\cl}\,\mathcal{D}'\phi^{\cl}\; W\big(\phi_0^{\cl},\pi_0^{\cl};t_0\big) \delta\Big( -\ddot{\phi}^{\cl} + \nabla^2 \phi^{\cl} -V^{(1)}\Big)~,
\label{eq:classical}
\end{align}
where the delta function is understood in the functional sense. We see that the CS approximation to the partition function makes use of the critical points only, i.e., the classical trajectories, with their initializations distributed according to the initial Wigner function.  Notice that the whole distribution function above is non-negative if the Wigner function is non-negative.

In comparison to the original expression~\eqref{eq:Z}, we may regard the $\phi^{\q}$ as hidden variables, only this time, the PDF is complex. However, this analogy is not quite complete, as we may also wish to measure $\phi^{\q}$-dependent operators. We want to stress that the complex PDF is a necessary condition for the violation of Bell inequalities, since if the distribution function is non-negative, one can always do the sampling and the generated samples cannot violate Bell inequalities. Now, with Eq.~\eqref{eq:classical}, we might speculate that there should not exist any violation of Bell inequalities in the free theory when a non-negative PDF about $\phi^{\cl}$ can be drawn. As we will see, this is not the case. 

%%%%%%%%%%%%%%%%%%%%%%%%%%%%%%%%%%%%%%%%%%%%%%%%%%%%%

\section{(Temporal) Leggett-Garg inequalities}
\label{sec:LG}

With regard the CS approximation, it will turn out that the temporal Bell-type inequalities due to Leggett and Garg~\cite{Leggett:1985zz} are of a richer structure, and we therefore choose to treat these before the more familiar spatial Bell inequalities.

The Leggett-Garg inequalities deal with measurements at different times. For the measurement operator, we choose $\hat Q={\rm sign}(\hat\phi)$, which is a proper dynamical operator that maps the continuous variable $\phi$ into a dichotomous one~\cite{Revzen, Martin:2017zxs}, taking values $\pm1$.

It is useful to consider which correlators the experiment can measure and which two-point correlation functions they correspond to in the theory. In an experiment, we prepare an ensemble of sets, consistent with the same initial state $|\psi\rangle$. For each set, a measurement $Q_1\equiv\braket{\hat{Q}_1}=r$ is read out at $t_1$ and another measurement $Q_2\equiv\braket{\hat{Q}_2}=s$ is read out at a later time $t_2$, with $r,s=\pm$. The joint probability $P(r,s)$ can then be calculated~\cite{Fritz}, e.g., as
\begin{align}
P(+,+) =\frac{N(+,+)}{\sum_{r,s=\pm}N(r,s)}~,
\end{align}
where $N(r,s)$ is the number of sets of $Q_1=r$ and $Q_2=s$.
Accordingly, the correlator is defined as
\begin{align}
C\equiv \sum_{r,s=\pm} rsP(r,s)~.
\label{eq:C}
\end{align}
For dichotomous variables, the correlator can further be related to the quantum two-point correlation function \cite{Fritz}. 
To see this, recall that we have two probabilities for the two measurements:
\begin{align}
P(r)=|\langle r;t_1|\psi\rangle|^2
\qquad \text{and} \qquad
P(s|r)=|\langle s;t_2|r;t_1\rangle|^2~,
\end{align}
which correspond to the probability of finding $r$ at $t_1$, given the initial state $|\psi\rangle$, and the probability of finding $s$ at $t_2$, given the state $|r;t_1\rangle$, i.e., the state into which the first measurement collapses the system.
The joint probability of finding $r$ at $t_1$ and $s$ at $t_2$ is then simply the product
\begin{equation}
P(r,s)=|\langle s;t_2|r;t_1\rangle|^2|  \langle r;t_1|\psi\rangle|^2
	=\langle\psi|r;t_1\rangle \langle r;t_1|s;t_2\rangle \langle s;t_2|r;t_1\rangle \langle r;t_1|\psi\rangle~.
\end{equation}
A dichotomous operator admits the following properties
\begin{align}
\hat Q&=\sum_{s=\pm1}s|s\rangle\langle s|,\qquad\frac{1+s\hat Q}{2}=|s\rangle\langle s|~,
\end{align}
and we can therefore write
\begin{align}
P(r,s)	&=\langle\psi|\frac{1+r\hat Q_1}{2} \frac{1+s\hat Q_2}{2} \frac{1+r\hat Q_1}{2} |\psi\rangle\nonumber\\&
	=\frac{1}{8}\langle\psi|\left(2+2r\hat Q_1+s\hat Q_2+rs\{\hat Q_1,\hat Q_2\}+s\hat Q_1\hat Q_2\hat Q_1\right)|\psi\rangle~.
\end{align}
It is then straightforward to obtain the following equalities:
\begin{subequations}
\begin{align}
\sum_{r,s=\pm1}P(r,s)&=1~,\\
\sum_{r,s=\pm1}rP(r,s)&=\langle\psi|\hat Q_1|\psi\rangle~,\\
\sum_{r,s=\pm1}sP(r,s)&=\frac{1}{2}\langle\psi|\hat Q_2+\hat Q_1\hat Q_2\hat Q_1|\psi\rangle~,\\
\sum_{r,s=\pm1}rsP(r,s)&=\frac{1}{2}\langle\psi|\{\hat Q_1,\hat Q_2\}|\psi\rangle~.
\label{eq:anticom}
\end{align}
\end{subequations}
The last of these [Eq.~\eqref{eq:anticom}] in particular means that the correlator measured in the experiment via Eq.~\eqref{eq:C} is really the quantum two-point correlation function involving the anticommutator of the measurement operators. Interestingly, the above four equations indicate that the measurement procedure allows us to find $\langle\psi|\hat Q_1|\psi\rangle$ and $\langle \psi| \{\hat Q_1, \hat Q_2\}  |\psi\rangle$, but not $\langle\psi|\hat Q_2|\psi\rangle$. As we shall see, this is not the case in the CS approximation, where we are able to find $\langle\psi|\hat Q_1|\psi\rangle$ and $\langle\psi|\hat Q_2|\psi\rangle$, but not $\langle \psi| \{\hat Q_1, \hat Q_2\}  |\psi\rangle$.

%%%%%%%%%%%%%%%%%%%%%%%%%%%%%%%%%%%%%%%%%%%%%%%%%%%%%

\subsection{Violation of the Leggett-Garg inequalities}

Since, in the case of the Leggett-Garg inequalities, the measurements are performed at the same spatial site, we can simplify the analysis significantly and consider the question in $d=0$ spatial dimensions, i.e., in QM.

As a concrete example, we focus on the quantum harmonic oscillator, subject to the Schr\"{o}dinger equation
\begin{align}
i\hbar \frac{\partial}{\partial t} \Psi( \phi;t) = \left[-\frac{\hbar^2}{2}\frac{\partial^2}{\partial  \phi^2} + \frac{\omega^2}{2} \phi^2 \right] \Psi( \phi;t)~.
\end{align}
We assume a Guassian initial state, displaced from the minimum of the potential well by an amount $\Delta$, i.e.,
\begin{align}
\Psi( \phi;t=0) = \left(\frac{\omega}{\hbar\pi}\right)^{1/4}e^{-\frac{\omega}{2\hbar}( \phi-\Delta)^2}~.
\end{align}
This leads to a positive Wigner function of the initial density matrix (with the explicit form shown in Eq.~\eqref{eq:Green_Wigner_free_QM}).
The full expression for the time-dependent wavefunction can be written in terms of the Feynman kernel
\begin{align}
&K( \phi,t; \phi',t')=
\langle  \phi,t |  \phi',t'\rangle
\nonumber \\
&\quad=\left\{\frac{\omega\csc[\omega(t-t')]}{2\pi i \hbar }\right\}^{1/2}
\exp\left(\frac{i\omega}{\hbar}\left\{\frac{\phi^2+ \phi'^2}{2}\cot[\omega (t-t')]-\phi\phi'\csc[\omega (t-t')]\right\}\right)~,
\end{align}
 as
\begin{align}
&\Psi( \phi;t)=\int \ud  \phi_0\; K( \phi,t; \phi_0,0) \Psi( \phi_0;0)~,
\end{align}
yielding
\begin{align}
\Psi(\phi,t) &= \left(\frac{\omega }{\hbar \pi}\right)^{1/4} \exp\left\{-\frac{\omega}{2\hbar}\left[\phi-\Delta \cos(\omega t)\right]^2-\frac{i\omega t}{2}-\frac{i\omega\Delta}{2\hbar}\sin(\omega t)\left[2\phi-\Delta\cos(\omega t)\right]\right\}~.
\end{align}
By inspection, we see that this corresponds to a Gaussian probability distribution, whose central peak oscillates about the origin with frequency $\omega$ and amplitude $\Delta$.
In fact, it is a coherent state of the form \cite{Klauder}
\begin{align}
\Psi_\alpha(\phi,t) &= \left(\frac{\omega }{\hbar \pi}\right)^{1/4}\exp\left\{-\frac{\omega}{2\hbar}\left[\phi-\sqrt{\frac{2\hbar}{\omega}}\,\mathrm{Re}\,\alpha(t)\right]^2+i\sqrt{\frac{2\omega}{\hbar}}\phi\, \mathrm{Im}\,\alpha(t)+i\theta(t)\right\}~,
\end{align}
with
\begin{equation}
\alpha(t)=\sqrt{\frac{\omega\Delta^2}{2\hbar}}\,e^{-i\omega t}\qquad \text{and}\qquad \theta(t)=-\frac{\omega t}{2}+\frac{1}{2}|\alpha(0)|^2\sin(2\omega t-\pi)~.
\end{equation}
We know the one-point function of the coherent state will oscillate between $\Delta$ and $-\Delta$, and this provides a perfect scenario for Leggett-Garg's proposal~\cite{Bose}.

With this in place, we can calculate the two-point function directly via
\begin{align}
\frac{1}{2}\langle\{ \hat Q_1, \hat Q_2\} \rangle  =\frac{1}{2} \left[\int \ud  \phi_1\, \ud  \phi_2\; \Psi^\dagger( \phi_2,t_2) Q(\phi_2) K( \phi_2,t_2; \phi_1,t_1) Q(\phi_1) \Psi(\phi_1,t_1) \right]+\text{c.c.}~,
\label{eq:sch_qm}
\end{align}
where $Q(\phi)\equiv {\rm sign}(\phi)$. This integral can be computed numerically, and we present some additional details of its structure in Appendix~\ref{App:integral}.

As an example, we can take $\omega=1$, $\hbar=1$ and $\Delta=2$, performing measurements at times $t_1=2.0$, $t_2=14.0$ and $t_3=19.6$.  This yields
\begin{equation}
C_{12} = 0.0407298~,\qquad C_{23} = 0.379699\qquad \text{and}\qquad  C_{13} = -0.680874~,
\end{equation}
and we find that the Leggett-Garg inequality is violated; namely,
\begin{equation}
C_{12}+C_{23}-C_{13}=1.1013 >1~,
\end{equation}
cf.~Eq.~\eqref{eq:genineq}.

%%%%%%%%%%%%%%%%%%%%%%%%%%%%%%%%%%%%%%%%%%%%%%%%%%%%%

\subsection{A conundrum}

As we argued earlier, the CS approximation cannot violate Bell inequalities, due to the non-negative PDF. We also saw that the CS approximation is actually an exact description of the free theory, for which we have just found a violation of Bell inequalities. So why do we get this apparently contradictory conclusion for the Bell inequalities?

\begin{figure}[!t]
\centering
\includegraphics[width=0.65\textwidth]{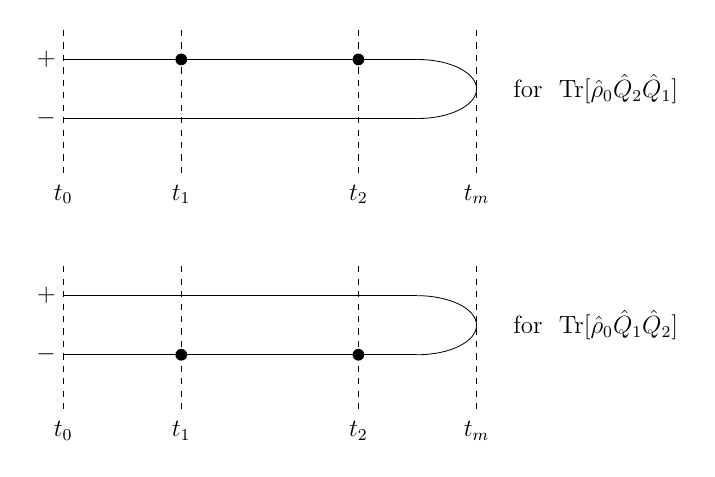}
\caption{\label{fig:ct_path} A plot showing the placement of operators on the closed-time contour for the terms needed in the anticommutator, with $t_2>t_1$ and $t_m$ being the latest time on the contour.}
\end{figure}

To answer this question, it is useful to check the definition of the QFT two-point function in the $\phi^{\cl}$-$\phi^{\q}$ basis; given $t_2 > t_1$,
\begin{subequations}
\begin{align}
C^{\rm QFT}(t_2,\mathbf{x}_2;t_1,\mathbf{x}_1)&=\frac{1}{2}\Big\langle \{{\rm sign}(\hat\phi(t_2,\mathbf{x}_2)),\;  {\rm sign}(\hat\phi(t_1,\mathbf{x}_1)) \}\Big\rangle
\nonumber\\
 &=\label{eq:C_QFT_pm}
\frac{
\int {\mathcal D}\phi\; W e^{-\AI}\, \frac{1}{2}\left[{\rm sign}(\phi^{+}_2) {\rm sign}(\phi^+_1) +{\rm sign}(\phi^-_1){\rm sign}(\phi^{-}_2)\right]
}{
\int {\mathcal D}\phi\; W e^{-\AI}
}\\
 &=\label{eq:C_QFT_q_cl}
\frac{
\int {\mathcal D}\phi\; W e^{-\AI}\, \frac{1}{2}\,{\rm sign}(\phi^{\cl}_2)\left[{\rm sign}(\phi^+_1) +{\rm sign}(\phi^-_1)\right]
}{
\int {\mathcal D}\phi\; W e^{-\AI}
} 
~,
\end{align}
\end{subequations}
where we have introduced the shorthand notation $\phi_i\equiv \phi(t_i,\mathbf{x}_i)$. We recall that $\mathcal{D}\phi\equiv \mathcal{D}\phi^+\mathcal{D}\phi^-$. The appearances of $\phi^\pm$ in Eq.~\eqref{eq:C_QFT_pm} can be understood by looking at Fig.~\ref{fig:ct_path}. Specifically, if we want to compute the correlator \smash{$\langle \hat Q_2\hat Q_1\rangle = {\rm Tr}\big[\hat{\rho}(t_0)\hat Q_2\hat Q_1\big]/{\rm Tr}\big[\hat{\rho}(t_0)\big]$}, we construct the path integral by inserting complete sets of states from $t_0$ to $t_1$, then from $t_1$ to $t_2$ and finally from $t_2$ back to $t_0$, which is just the path given in the upper plot of Fig.~\ref{fig:ct_path}. The operator at the larger time may appear on either branch of the contour without affecting the result. In fact, we can contract the path so that it ends at the larger time, i.e., $t_2$, and then use the results of Refs.~\cite{Mou:2019tck, Mou:2019gyl} to set $\phi^{\q}(t_2,x)=0$, yielding the final line of Eq.~\eqref{eq:C_QFT_q_cl}.

Instead, the CS approximation can only compute (if we restore both the $\phi^+$ and $\phi^-$ integrals),
\begin{align}
&C^{\rm CS}(t_2,\mathbf{x}_2;t_1,\mathbf{x}_1)=\frac{
\int {\mathcal D}\phi\;W e^{-\AI}\, {\rm sign}(\phi^{\cl}_2) {\rm sign}(\phi^{\cl}_1)
}{
\int {\mathcal D}\phi\;W e^{-\AI}
}~.
\end{align}
We can already conclude that $C^{\rm QFT}$ and $C^{\rm CS}$ do not calculate the same thing, and that the temporal Bell inequalities rely upon more than just $\phi^{\cl}$.

For the free scalar field, we can proceed further, by integrating out $\phi^{\q}$. This can be achieved by first discretising the path integral, as we describe in detail in Appendix~\ref{sec:appendix_a}. We then find
\begin{align}
C^{\rm QFT}(t_2,\mathbf{x}_2;t_1,\mathbf{x}_1)
=
\frac{ 
\int {\mathcal D} \phi^{\cl}_0\, {\mathcal D} \pi^{\cl}_0\,\ud \pi^{\cl}_1\;W
 ~{\rm sign}(\tilde{\tilde{\phi}}^{\cl}_2)\frac{1}{\pi}
 \frac{
\sin\left(\ud^d\mathbf{x}\frac{2\tilde\phi^{\cl}_1}{\hbar }\left(\pi^{\cl}_1-\tilde\pi^{\cl}_1\right)\right)}
{\pi^{\cl}_1-\tilde\pi^{\cl}_1} 
}{
\int  {\mathcal D} \phi^{\cl}_0\,  {\mathcal D} \pi^{\cl}_0\;W
} 
~,
\label{eq:QFTnew}
\end{align}
in its discrete form, with the tilded and double-tilded variables defined below.  The direct appearance of the infinitesimal $\ud^d \mathbf{x}$ reflects the difference between the functional derivative $\delta \phi/\delta \pi$ and the normal derivative $\partial \phi/\partial \pi$. The final result, however, will not involve the infinitesimal [see Eq.~\eqref{eq:qft} below].

The variable $\tilde \phi^{\cl}(t,\mathbf{x})$ denotes the field as evolved from the initial data $\tilde \phi^{\cl}(t_0,\mathbf{x})=\phi_0^{\cl}(\mathbf{x})$ and $\dot{\tilde \phi}^{\cl}(t_0,\mathbf{x})=\pi_0^{\cl}(\mathbf{x})$ via the classical equation of motion, which may be represented using the retarded Green function $G_R$ as
\begin{gather}
\tilde \phi^{\cl}(t,\mathbf{x}) = \int \ud^d \mathbf{x}_0\;\left[G_R(t,\mathbf{x};t_0,\mathbf{x}_0)\pi^{\cl}_0(\mathbf{x}_0)-\frac{\partial G_R(t,\mathbf{x};t_0,\mathbf{x}_0)}{\partial t_0}\,\phi_0^{\cl}(\mathbf{x}_0)\right]
~.
\end{gather}
The associated momentum field is defined as
\begin{equation}
\tilde \pi^{\cl}(t,\mathbf{x})\equiv \lim_{\ud t\to 0}\frac{\tilde \phi^{\cl}(t+\ud t,\mathbf{x})-\tilde \phi^{\cl}(t,\mathbf{x})}{\ud t}~.
\end{equation}
There are two more fields involved in the expression: The integration variable
\begin{equation}
\pi^{\cl}(t_1,\mathbf{x}_1)\equiv  \lim_{\ud t\to 0}\frac{\phi^{\cl}(t_1+\ud t,\mathbf{x}_1)-\tilde \phi^{\cl}(t_1,\mathbf{x}_1)}{\ud t}
\end{equation}
(or $\pi^{\cl}_1$ for the shorthand notation) is the momentum field at time $t_1$ and position $\mathbf{x}_1$, and the field $\tilde{\tilde{\phi}}^{\cl}_2$ depends on $\pi^{\cl}_1$ via
\begin{align}
\tilde{\tilde{\phi}}^{\cl}_2 =& \int \ud^d \mathbf{x}_1\;\left[G_R(t_2,\mathbf{x}_2;t_1,\mathbf{x}_1)\tilde\pi^{\cl}_1  - \frac{\partial G_R(t_2,\mathbf{x}_2;t_1,\mathbf{x}_1)}{\partial t_1}\, \tilde\phi^{\cl}_1\right]
\nonumber\\\nonumber & \quad
+\ud^d \mathbf{x}\,G_R(t_2,\mathbf{x}_2;t_1,\mathbf{x}_1)\left[\pi^{\cl}_1 - \tilde\pi^{\cl}_1 \right] 
\\
=&\tilde \phi^{\cl}_2 +\ud^d \mathbf{x}\; G_R(t_2,\mathbf{x}_2;t_1,\mathbf{x}_1)\left[\pi^{\cl}_1 - \tilde\pi^{\cl}_1 \right]~.
\label{eq:ddx}
\end{align}
That is, $\tilde{\tilde{\phi}}^{\cl}_2$ is almost a classical solution, except with the momentum coming from time $t_1$ and position $\mathbf{x}_1$, which is determined by the integration variable $\pi^{\cl}_1$.
Eventually, the integration over $\pi^{\cl}_1$ leads to
\begin{align}
C^{\rm QFT}=
\frac{ 
\int {\mathcal D} \phi^{\cl}_0\, {\mathcal D} \pi^{\cl}_0\;W\, 
\frac{2}{\pi}\,{\rm Si}\left(\frac{2\tilde\phi^{\cl}_2\tilde\phi^{\cl}_1}{\hbar |G_R(t_2,\mathbf{x}_2;t_1,\mathbf{x}_1)|}\right)
}{
\int  {\mathcal D} \phi^{\cl}_0\,  {\mathcal D} \pi^{\cl}_0\; W
}~,
\label{eq:qft}
\end{align}
where \smash{${\rm Si}(w)\equiv \int_0^{w}\ud u \sin(u)/u$} is the sine integral function, and we have made use of the identity \smash{$\int_{-\infty}^{+\infty}\ud u~ {\rm sign}(a+bu)\sin(cu)/u=2\int_{0}^{a/|b|}\ud u \sin(cu)/u$}.
For comparison, the CS approximation leads to
\begin{align}
C^{\rm CS}=
\frac{ 
\int {\mathcal D}\phi^{\cl}_0\, {\mathcal D}\pi^{\cl}_0\; W  {\rm sign}(\tilde\phi^{\cl}_2)  {\rm sign}(\tilde\phi^{\cl}_1)
}{
\int {\mathcal D}\phi^{\cl}_0\, {\mathcal D}\pi^{\cl}_0\; W  
}~.
\label{eq:cs}
\end{align}

Returning to our QM example, the retarded Green's function, for $t_2>t_1$,
and the normalized Wigner function are given respectively by
\begin{align}\label{eq:Green_Wigner_free_QM}
\hspace{-0.2cm}
G_R(t_2;t_1)=\frac{\sin\left[\omega (t_2-t_1)\right]}{\omega} \quad \text{and} \quad
W\big(\phi_0^{\cl},\pi_0^{\cl};t_0\big) = \frac{1}{\pi\hbar}
\exp\left[-\frac{\omega}{\hbar}\big(\phi_0^{\cl}-\Delta\big)^2 -\frac{(\pi_0^{\cl})^2}{\hbar \omega} \right]~.
\end{align}
The two-point function \eqref{eq:qft} is then
\begin{align}
C^{\rm QFT}&=
\frac{2}{\hbar\pi^2}
\int \ud \phi_0^{\cl}\, \ud \pi_0^{\cl}\;\exp\left[-\frac{\omega}{\hbar}\big(\phi_0^{\cl}-\Delta\big)^2 -\frac{(\pi_0^{\cl})^2}{\hbar \omega} \right]
\nonumber\\
&\phantom{=}\times
{\rm Si}\left\{\frac{2 \omega\left[\phi_0^{\cl}\cos(\omega t_1)+\pi_0^{\cl}\frac{\sin(\omega t_1)}{\omega}\right] \left[\phi_0^{\cl}\cos(\omega t_2)+\pi_0^{\cl}\frac{\sin(\omega t_2)}{\omega}\right]}{\hbar |\sin[\omega(t_2-t_1)]|}\right\}~.
\end{align}
Less obvious is that this expression yields the same result as that given by Eq.~\eqref{eq:sch_qm}, which one can evaluate numerically.

Notice that we could obtain the correct violation of the Bell inequality from a CS calculation if we were to use the ${\rm Si}$ function and retarded Green's function as in Eq.~\eqref{eq:qft}, instead of the product of ${\rm sign}$ functions that the CS approximation naively leads to. This prescription, however, only applies to the free theory, and there is no obvious way to determine what replacement should be used for general interactions.

%%%%%%%%%%%%%%%%%%%%%%%%%%%%%%%%%%%%%%%%%%%%%%%%%%%%%

\section{(Spatial) Bell inequalities}
\label{sec:Bell}

An interesting observation is that the ${\rm Si}$ function appearing in Eq.~\eqref{eq:qft} admits the following limit:
\begin{align}
\lim_{|G_R|\to 0}\frac{2}{\pi}\,{\rm Si}\left(\frac{2\tilde\phi^{\cl}_2\tilde\phi^{\cl}_1}{\hbar |G_R(t_2,\mathbf{x}_2;t_1,\mathbf{x}_1)|}\right)
={\rm sign}(\tilde\phi^{\cl}_2)  {\rm sign}(\tilde\phi^{\cl}_1)~.
\end{align}
Since the retarded propagator $G_R(t_2,\mathbf{x}_2;t_1,\mathbf{x}_1)$ is causal, it vanishes when the separation of two points is spacelike, i.e., $(t_2-t_1)^2<|\mathbf{x}_2-\mathbf{x}_1|^2$, which means that, for the free theory, $C^{\rm QFT}$ and $C^{\rm CS}$ give the same result in the case of spatial Bell inequalities.
On the other hand, when the Wigner function is non-negative, the two-point functions in the form of Eq.~\eqref{eq:cs} can never violate Bell inequalities. Thus, there can be no violation of Bell inequalities among spacelike correlation functions in free scalar field QFT, unless some entanglement exists in the initialization. 

Beyond the free theory, we point out that it is a general property of QFT that $\phi^{\q}$ does not appear explicitly in the calculation of the spatial Bell inequalities.
In fact, the following equations are valid in general in QFT:
\begin{align}
\label{eq:CFTgeneral}
C^{\rm QFT}(t_2,\mathbf{x}_2;t_1,\mathbf{x}_1)\big|_{\rm spacelike}&=\frac{1}{2}\Big\langle \{{\rm sign}(\hat\phi_2),\;  {\rm sign}(\hat\phi_1) \}\Big\rangle
\nonumber\\
 &=
\frac{
\int {\mathcal D}\phi\; W e^{-\AI}\, \frac{1}{2}\left[{\rm sign}(\phi^{+}_2) {\rm sign}(\phi^+_1) +{\rm sign}(\phi^-_1){\rm sign}(\phi^{-}_2)\right]
}{
\int {\mathcal D}\phi\; W e^{-\AI}
} 
\nonumber\\
 &=
\frac{
\int {\mathcal D}\phi\; W e^{-\AI}\, {\rm sign}(\phi^{\cl}_2) {\rm sign}(\phi^{\cl}_1)
}{
\int {\mathcal D}\phi\; W e^{-\AI}}~,
\end{align}
with the reason discussed in Appendix~\ref{sec:appendix_a}.
In comparison, the CS approximation computes 
\begin{align}
\label{eq:CSgeneral}
C^{\rm CS}(t_2,\mathbf{x}_2;t_1,\mathbf{x}_1)&=
\frac{
\int {\mathcal D}\phi\; W e^{-\AI'}\, {\rm sign}(\phi^{\cl}_2) {\rm sign}(\phi^{\cl}_1)
}{
\int {\mathcal D}\phi\; W e^{-\AI'}}~,
\end{align}
where $\AI'$ denotes the $\AI$ with the higher-order terms in $\phi^{\q}$ discarded.
In what follows, we will refer to those higher-order terms in $\phi^{\q}$ as ``quantum vertices'', a name that can reflect their roles in Feynman diagrams \cite{Aarts:1997kp,Epelbaum:2014yja,Mou:2019tck}.
Since the CS approximation with Eq.~\eqref{eq:CSgeneral} cannot violate any Bell-like inequalities, while the full quantum theory with Eq.~\eqref{eq:CFTgeneral} can, and given that the only difference between the two concerns the quantum vertices, we might speculate that these quantum vertices must have something to do with the origin of quantum entanglement.
(Here, we equate quantum entanglement with the violation of Bell inequalities.)
This seems to be the case with the following arguments.

We first notice that there exists a loophole in the above reasoning, which is related to the initialization.
When quantum entanglement appears in the initial state, the Wigner function will be  negative-valued in some field region.
The negative distribution will make the standard sampling method difficult, if not impossible, to generate initializations for the CS approximation.
That being said, if there exists some sophisticated re-sampling method for doing the initialization, the CS approximation will escape the constraint of not violating Bell inequalities.

On the other hand, the entanglement that already exists in the initialization does not help us to understand the origin of quantum entanglement.
For this purpose, the best scenario is where the system starts from a non-negative Wigner function and gains some violation of spatial Bell inequalities during the evolution.
Consider, for instance, the decay of a spinless particle into two photons. 
In this case, as we argued above, the CS, as an approximation, cannot capture such quantum entanglement.
We can therefore conclude that it is those higher-order $\phi^{\q}$-terms in the action\,---\,the quantum vertices\,---\,that make quantum entanglement possible, and it is the absence of them that renders the CS approximation unable to capture quantum entanglement.

We further point out in the $\phi^{\cl}$-$\phi^{\q}$ representation that the quantum properties are usually accompanied by the appearance of $\phi^{\q}$.
Recall, from the last section, the central role that the $\phi^{\q}$ plays in the Leggett-Garg or temporal Bell inequalities and here in the origin of the violation of spatial Bell inequalities, through the quantum vertices.

Another example is the commutation relation,
\begin{align}
\langle [\hat\phi(t,\mathbf{x}),\hat\pi(t,\mathbf{y})] \rangle = i\hbar \delta^d\left(\mathbf{x}-\mathbf{y}\right)~,
\end{align}
which is in fact a $\q$-$\cl$ two-point function. To see this, first recall, from the field-configuration representation of the path integral, that the field momentum can be computed via
\begin{align}
\pi(t,\mathbf{y})\equiv \lim_{\ud t\to 0} \frac{\phi(t+\ud t,\mathbf{y})-\phi(t,\mathbf{y})}{\ud t}~.
\end{align}
It is then straightforward to derive the commutation relation and obtain
\begin{subequations}
\begin{align}
\langle [\hat\phi(t,\mathbf{x}),\hat\pi(t,\mathbf{y})] \rangle
=& \lim_{\ud t\to 0}\frac{\langle \hat\phi(t,\mathbf{x})\hat\phi(t+\ud t,\mathbf{y})-  \hat\phi(t+\ud t,\mathbf{y})\hat\phi(t,\mathbf{x})\rangle}{\ud t}
\nonumber
\\
=& \lim_{\ud t\to 0}\frac{\langle \phi^-(t,\mathbf{x})\phi^-(t+\ud t,\mathbf{y})-  \phi^+(t+\ud t,\mathbf{y})\phi^+(t,\mathbf{x})\rangle}{\ud t}
\nonumber
\\
\label{eq:comm1}
=& -\lim_{\ud t\to 0}\frac{\langle \phi^{\q}(t,\mathbf{x})\phi^{\cl}(t+\ud t,\mathbf{y})\rangle}{\ud t}
\\
\label{eq:comm2}
=& -\langle \phi^{\q}(t,\mathbf{x})\pi^{\cl}(t,\mathbf{y})\rangle
~.
\end{align}
\end{subequations}
In the free theory, the relevant two-point function in Eq.~\eqref{eq:comm1} is given by
\begin{align}
\langle \phi^{\q}(t_1,\mathbf{x})\phi^{\cl}(t_2,\mathbf{y}) \rangle
=-i\hbar\theta\left(t_2-t_1\right)
\int\frac{\ud^d \mathbf{p}}{(2\pi)^d}\frac{\sin\left(\omega_\mathbf{p}(t_2-t_1)\right)}{\omega_\mathbf{p}}e^{i\mathbf{p}\cdot(\mathbf{x}-\mathbf{y})}~,
\end{align}
which, as a double-check, gives the correct delta function in the limit $\ud t\to 0$.

In retrospect, when facing Eq.~\eqref{eq:Green_Wigner_free_QM}, the explicit Wigner function in the example, and noticing that $\phi_0^{\cl}$ and $\pi_0^{\cl}$ are independently distributed, one might doubt whether the initialization would respect the commutation relation.
Now, as we see, one should in fact compute the two-point function in Eq.~\eqref{eq:comm2} in order to check the commutation relation.
Note that $\phi_0^q$ and $\pi_0^{\cl}$ appear in the kernel of the Weyl transform \eqref{eq:Weyl}.
Thus, by a substitution $\phi_0^{\q} W = \left(i\hbar \delta W/\delta \pi^{\cl}_0\right)$ and then integration by parts, we can verify that $\langle [\hat\phi(t_0,\mathbf{x}),\hat\pi(t_0,\mathbf{y})] \rangle=i\hbar \delta^d(\mathbf{x}-\mathbf{y})$.
In particular, the validity does not depend on the detail of the Wigner function or the initial density matrix, but only on the Weyl transform.

%%%%%%%%%%%%%%%%%%%%%%%%%%%%%%%%%%%%%%%%%%%%%%%%%%%%%

\section{Conclusions}
\label{sec:conc}

We have studied QFT from the viewpoint of statistics. A general feature of the real-time path integrals is that the PDF is complex-valued, and, as a result, standard statistical tools, such as Markov chain or Monte Carlo, cannot be applied directly. This is reminiscent of the so-called ``numerical sign problem''. To deal with the complex path integral, we can seek to use the Lefschetz thimble method, and we can summarize the following general properties of the real-time path integral (for further details, see Refs.~\cite{Mou:2019tck, Mou:2019gyl}):

\begin{itemize}

\item
The real-time path integral admits a two-part separation into the initial density matrix (via the Wigner function $W$) and the dynamical part ($e^{-\AI}$). The initial Wigner function provides the initial data for the dynamical part, and it can be either non-negative or generally real-valued. The dynamical part, on the other hand, is purely a phase term, which is always situated on the edge of the convergent region in the complexified $\phi$-space.

\item
Applying the Lefschetz thimble approach to the dynamical part, we find that the saddle points consist of classical trajectories. In particular, for each initialization, there exists a unique classical trajectory/saddle point/Lefschetz thimble.

\item
The exponent in the dynamical part ($e^{-\AI}$) includes only odd terms in $\phi^{\q}$. If we throw away any higher-order terms in $\phi^{\q}$, integrating out the  $\phi^{\q}$ leads to functional delta functions that pick out the classical solutions. This leads to the CS approximation. Note that, after they have been integrated out, the $\phi^{\q}$ fields become {\it hidden} in the sense that we can no longer compute $\phi^{\q}$-dependent operators within the CS approximation. 

\end{itemize}

In this paper, we have further pointed out that the complex PDF is a necessary condition for the violation of Bell-type inequalities, since, if the distribution function is non-negative, one can always apply the sampling method, and the so-generated samples cannot violate Bell-type inequalities. In this sense, the CS approximation is not expected to yield any violation of Bell-type inequalities, as it is restricted to non-negative PDFs.

This observation, however, leads to a conundrum that the free theory in a coherent state, for which the CS approach makes no approximations, admits a violation of the Leggett-Garg or temporal Bell inequalities.
We resolved this puzzle by demonstrating that, for the example measurement operator $\hat Q={\rm sign}(\hat\phi)$, the temporal two-point function depends on both $\phi^{\cl}$ \emph{and} $\phi^{\q}$. Note that the PDF involving $\phi^{\cl}$ only is non-negative, but the PDF with $\phi^{\q}$ included is complex. This further validates our conclusion above that a complex PDF is needed for violations of Bell-type inequalities.

We have also identified a key difference between the spatial and temporal Bell-type inequalities. We first recall that, in any (local) QFT~\cite{Headrick:2019eth}, any two operators with spacetime arguments that are spacelike separated commute. In the language of $\phi^{\cl}$-$\phi^{\q}$, this is obvious because the product of two operators contains $\phi^{\cl}$ only. Thus, the spatial Bell inequalities have classical analogues, and one can therefore compute them directly within the CS approximation (but one will not see a violation).
In comparison, there are no such analogues for the temporal Bell-type inequalities due to the explicit dependence on $\phi^{\q}$. In this case, the CS approximation should not be used to compute Leggett-Garg inequalities directly, not even as an approximation.

These observations in relation to the CS approximation and Bell-type inequalities can be summarized as follows:
\begin{enumerate}

\item The CS approximation is consistent with the full QFT treatment for a non-negative initial Wigner function in the case of spatial Bell inequalities and in the absence of interactions, because the configuration $\phi^{\q}=0$ is picked out.

\item The CS approximation fails to capture violations of spatial Bell inequalities for the free theory, because we are restricted to non-negative Wigner functions.

\item We suspect that the CS approximation cannot be made consistent for spatial Bell inequalities for interacting theories, even for a non-negative initial Wigner function, because of the non-trivial $\phi^{\q}$ dependence, which will prevent us from identifying a CS equivalent for the true QFT correlation function.

\item The naive application of the CS approximation to the free theory cannot capture the violation of temporal Bell inequalities due to the $\phi^{\q}$ dependence of the measurement operators.  While, for the free theory, we have been able to identify a CS-equivalent correlation function (viz.~the sine integral expression rather than the product of sign functions in our example), the identification of such a prescription may not be straightforward in the interacting case, as per point 3.~above.

\end{enumerate}
The CS approximation is quantitatively sound when occupation numbers are large, i.e., $\phi^{\cl}$ is much larger than $\phi^{\q}$. Discarding higher-order terms in $\phi^{\q}$ altogether then allows straightforward sampling of a non-negative initial PDF and evolution with classical equations of motion. While this prescription follows explicitly from the quantum path integral, as we have described, it amounts in practice to performing a (ensemble-averaged) classical field theory simulation. However, what we have seen in this work is that attempts to stretch the CS approximation to truly quantum phenomena require extreme care. In the non-interacting case, one would expect agreement between quantum and classical results because of the linearity of both the quantum operator equations and the classical equations of motion. However, we have seen that the CS prescription does not in general give the correct qualitative result, because the observables themselves may be constructed from the $\phi^{\q}$. As a result, the step in which $\phi^{\q}$ is integrated out does not go through. It is certainly illuminating to think of the CS approximation as a limit of the thimble formalism of the quantum path integral, but its applicability remains valid only for phenomena where $\phi^{\cl} \gg \phi^{\q}$, both for the evolution and the observables\,---\,the classical realm.

Having found that the CS prescription can fail even for a free, linear theory, one may wonder about quantum corrected or semiclassical approaches, such as (truncated) Schwinger-Dyson (SD)/Kadanoff-Baym (KB) evolution. These involve solving directly for the (quantum) correlators, the Greens functions, and so for a free theory the result must be exact, as it involves no approximation of the evolution nor (in principle) of the initial conditions. For instance, unlike the CS approximation, KB preserves the basic quantum commutator and hence the ``half'', zero-point fluctuations. From the point of view of computing Leggett-Garg inequalities, the issue becomes how to express the observable in terms of the $n$-point functions of the theory. This is non-trivial, and likely involves computing an entire series expansion. We have not attempted this here.

%%%%%%%%%%%%%%%%%%%%%%%%%%%%%%%%%%%%%%%%%%%%%%%%%%%%%

\section*{Acknowledgements}

The work of PM was supported by a Leverhulme Trust Research Leadership Award (Grant No.~RL-2016-028) and a Nottingham Research Fellowship from the University of Nottingham. The work of PS was supported in part by STFC grant ST/P000703/1.

%%%%%%%%%%%%%%%%%%%%%%%%%%%%%%%%%%%%%%%%%%%%%%%%%%%%%

\appendix

%%%%%%%%%%%%%%%%%%%%%%%%%%%%%%%%%%%%%%%%%%%%%%%%%%%%%

\section{QM correlation function}
\label{App:integral}

In order to make contact with the Schwinger-Keldysh calculation, we can write the correlation function [Eq.~\eqref{eq:sch_qm}] in the form
\begin{align}
\frac{1}{2}\langle\{ \hat Q_1, \hat Q_2\} \rangle  =\left[\frac{1}{2} \int \ud  \phi^+\, \ud  \phi^- \;\Psi^\dagger( \phi^-,t_2) Q( \phi_-) K( \phi^-,t_2; \phi^+,t_1) Q( \phi_+) \Psi( \phi^+,t_1)  \right]+\text{c.c.}~.
\end{align}
After making all the relevant substitutions, this becomes
\begin{align}
\frac{1}{2}\langle\{ \hat Q_1, \hat Q_2\} \rangle  &=\frac{\omega}{2 \pi\hbar}\left(1-e^{2i\omega(t_1-t_2)}\right)^{-1/2}e^{-\frac{\omega\Delta^2}{4\hbar}\left(2+e^{-2i\omega t_1}+e^{2i\omega t_2}\right)}\nonumber\\&\phantom{=}\times\left[-\int_{-\infty}^{+\infty}\frac{{\rm d}\xi_+}{\pi}\int_{-\infty}^{+\infty}\frac{{\rm d}\xi_-}{\pi}\;\frac{\xi_{+}}{\xi_{+}^2+\epsilon^2}\frac{\xi_{-}}{\xi_{-}^2+\epsilon^2}\right]
\nonumber\\&\phantom{=}\times\left[\int \ud  \phi_+\, \ud  \phi_-\; e^{-A[(\phi^+)^2+(\phi^-)^2]-B^+\phi^+-B^-\phi^--C \phi^+\phi^-}\right]+\text{c.c.}~,
\label{eq:intermediate}
\end{align}
where we have written
\begin{equation}
{\rm sign}(\phi^{\pm})=i\int_{-\infty}^{+\infty}\frac{{\rm d}\xi_{\pm}}{\pi}\frac{\xi_{\pm}}{\xi_{\pm}^2+\epsilon^2}e^{-i\phi^{\pm}\xi_{\pm}}~,
\end{equation}
with $\epsilon\equiv 0^+$, and defined
\begin{subequations}
\begin{gather}
A\equiv \frac{\omega}{2\hbar}\left(1-i\cot[\omega(t_2-t_1)]\right)~,\\
\quad B^+\equiv -\frac{\omega\Delta}{\hbar}e^{-i\omega t_1}+i\xi_+~,\quad B^-\equiv -\frac{\omega\Delta}{\hbar}e^{i\omega t_2}+i\xi_-~,\\
C\equiv \frac{i\omega}{\hbar}{\rm cosec}[\omega(t_2-t_1)]~.
\end{gather}
\end{subequations}
The $\phi^+$ and $\phi^-$ integrals can be performed analytically, so long as ${\rm Re}\left[4A^2-C^2\right]>0$, and we arrive at the result
\begin{align}
\frac{1}{2}\langle\{ \hat Q_1, \hat Q_2\} \rangle &=-\int_{-\infty}^{+\infty}\frac{{\rm d}\xi_{+}}{\pi}\int_{-\infty}^{+\infty}\frac{{\rm d}\xi_{-}}{\pi}\frac{\xi_{+}}{\xi_{+}^2+\epsilon^2}\frac{\xi_{-}}{\xi_{-}^2+\epsilon^2}\nonumber\\&\phantom{=}\times {\rm Re}\left\{e^{-\frac{\hbar}{4\omega}\left[\xi_+^2+2e^{i\omega(t_1-t_2)}\xi_+\xi_-+\xi_-^2\right]-i\Delta\left[\xi_+\cos(\omega t_1)+\xi_-\cos(\omega t_2)\right]}\right\}\nonumber\\
&=-\int_{-\infty}^{+\infty}\frac{{\rm d}\xi_{+}}{\pi}\int_{-\infty}^{+\infty}\frac{{\rm d}\xi_{-}}{\pi}\frac{\xi_{+}}{\xi_{+}^2+\epsilon^2}\frac{\xi_{-}}{\xi_{-}^2+\epsilon^2} e^{-\frac{\hbar}{4\omega}\left(\xi_+^2+2\cos[\omega(t_1-t_2)]\xi_+\xi_-+\xi_-^2\right)}\nonumber\\&\phantom{=}\times\cos\left\{\frac{\hbar}{2\omega}\sin[\omega(t_1-t_2)]\xi_+\xi_-+\Delta\left[\xi_+\cos(\omega t_1)+\xi_-\cos(\omega t_2)\right]\right\}~.
\end{align}
After making the change of variables
\begin{equation}
\sqrt{\hbar/\omega}\xi_{\pm}\to \xi_{\pm}~,\quad \sqrt{\hbar/\omega}\epsilon\to \epsilon~,
\end{equation}
we have
\begin{align}
\frac{1}{2}\langle\{ \hat Q_1, \hat Q_2\} \rangle &=-\int_{-\infty}^{+\infty}\frac{{\rm d}\xi_{+}}{\pi}\int_{-\infty}^{+\infty}\frac{{\rm d}\xi_{-}}{\pi}\frac{\xi_{+}}{\xi_{+}^2+\epsilon^2}\frac{\xi_{-}}{\xi_{-}^2+\epsilon^2} e^{-\frac{1}{4}\left(\xi_+^2+2\cos[\omega(t_1-t_2)]\xi_+\xi_-+\xi_-^2\right)}\nonumber\\&\phantom{=}\times\cos\left\{\frac{1}{2}\sin[\omega(t_1-t_2)]\xi_+\xi_-+\sqrt{\frac{\omega \Delta^2}{\hbar}}\left[\xi_+\cos(\omega t_1)+\xi_-\cos(\omega t_2)\right]\right\}~\nonumber\\&=-\int_{-\infty}^{+\infty}\frac{{\rm d}\xi_{+}}{\pi}\int_{-\infty}^{+\infty}\frac{{\rm d}\xi_{-}}{\pi}\frac{\xi_{+}}{\xi_{+}^2+\epsilon^2}\frac{\xi_{-}}{\xi_{-}^2+\epsilon^2} e^{-\frac{1}{4}\left(\xi_++\xi_-\right)^2+\sin^2[\omega(t_1-t_2)/2]\xi_+\xi_-}\nonumber\\&\phantom{=}\times\cos\left\{\frac{1}{2}\sin[\omega(t_1-t_2)]\xi_+\xi_-+\sqrt{\frac{\omega \Delta^2}{\hbar}}\left[\xi_+\cos(\omega t_1)+\xi_-\cos(\omega t_2)\right]\right\}~,
\end{align}
in which we see that the only dependence on the model parameters is through the parameter $c=\omega \Delta^2/(2\hbar)$.

Instead, by making the change of variables
\begin{equation}
\phi^{\pm}=\phi^{\cl}\pm\frac{1}{2}\phi^{\q}
\end{equation}
in Eq.~\eqref{eq:intermediate}, the integral can be written in the form
\begin{align}
\frac{1}{2}\langle\{ \hat Q_1, \hat Q_2\} \rangle  &=\frac{\omega}{2 \pi\hbar}\left(1-e^{2i\omega(t_1-t_2)}\right)^{-1/2}e^{-\frac{\omega\Delta^2}{4\hbar}\left(2+e^{-2i\omega t_1}+e^{2i\omega t_2}\right)}
\nonumber\\&\phantom{=}\times\left[-\int_{-\infty}^{+\infty}\frac{{\rm d}\xi_+}{\pi}\int_{-\infty}^{+\infty}\frac{{\rm d}\xi_-}{\pi}\;\frac{\xi_{+}}{\xi_{+}^2+\epsilon^2}\frac{\xi_{-}}{\xi_{-}^2+\epsilon^2}\right]\nonumber\\&\phantom{=}\times
\left[\int \ud  \phi^{\cl}\, \ud  \phi^{\q}\; e^{-A^{\cl}(\phi^{\cl})^2-B^{\cl}\phi^{\cl}-A^{\q}(\phi^{\q})^2-B^{\q}\phi^{\q}}\right]+\text{c.c.}~,
\end{align}
where
\begin{subequations}
\begin{align}
A^{\cl}&\equiv2A+C=\frac{\omega}{\hbar}\left(1+i \tan\left[\omega(t_2-t_1)/2\right]\right)~,\\
\quad A_{\rm q}&\equiv\frac{1}{4}(2A-C)=\frac{\omega}{4\hbar}\left(1-i \cot\left[\omega(t_2-t_1)/2\right]\right)~,\\
B^{\cl}&\equiv B^++B^-=-\frac{\omega \Delta}{\hbar}\left(e^{-i\omega t_1}+e^{i\omega t_2}\right)+i(\xi_++\xi_-)~,\\
B^{\q}&\equiv\frac{1}{2}\left(B^+-B^-\right)=-\frac{\omega \Delta}{2\hbar}\left(e^{-i\omega t_1}-e^{i\omega t_2}\right)+\frac{i}{2}(\xi_+-\xi_-)~.
\end{align}
\end{subequations}
The factor
\begin{equation}
\left(1-e^{2i\omega(t_1-t_2)}\right)^{-1/2}\int \ud  \phi^{\q}\; e^{-A^{\q}(\phi^{\q})^2-B^{\q}\phi^{\q}}\underset{t_1=t_2}{\longrightarrow} \sqrt{\frac{\pi h}{\omega}}
\end{equation}
is independent of time in the limit $t_1=t_2$. In fact, this is because one obtains a delta function of $\phi^{\q}$ in this limit. To see this, we need to make use of the following limit representation of the delta function:
\begin{equation}
\lim_{\varepsilon\to 0^+}\frac{1}{\sqrt{2\pi i \varepsilon}}e^{\frac{ix^2}{2\varepsilon}}=\delta(x)~,
\end{equation}
with
\begin{equation}
\varepsilon=\tan[\omega(t_2-t_1)/2]~,\quad x=\sqrt{\frac{\omega}{2\hbar}}\phi^{\q}~.
\end{equation}
We can then take
\begin{align}
&\lim_{t_1\to t_2}\left(1-e^{2i\omega(t_1-t_2)}\right)^{-1/2}\int \ud  \phi^{\q}\; e^{-A^{\q}(\phi^{\q})^2-B^{\q}\phi^{\q}}=\lim_{t_1\to t_2}\left(\frac{2\pi i\tan[\omega(t_2-t_1)/2]}{1-e^{2i\omega(t_1-t_2)}}\right)^{1/2}\nonumber\\&\phantom{=}\times\int{\rm d}\phi^{\q}\;e^{-\frac{\omega}{4\hbar}(\phi^{\q})^2-B^{\q}\phi^{\q}}\left(2\pi i\tan[\omega(t_2-t_1)/2]\right)^{-1/2}e^{i \frac{\omega}{4\hbar}(\phi^{\q})^2\cot[\omega(t_2-t_1)/2]}~,
\end{align}
and since
\begin{equation}
\lim_{t_1\to t_2}\left(\frac{2\pi i\tan[\omega(t_2-t_1)/2]}{1-e^{2i\omega(t_1-t_2)}}\right)^{1/2}=\sqrt{\frac{\pi}{2}}~,
\end{equation}
and
\begin{equation}
\lim_{t_1\to t_2}B^{\q}=\frac{i\omega \Delta}{\hbar}\sin[\omega t_2]+\frac{i}{2}(\xi_+-\xi_-)~,
\end{equation}
we are justified in writing
\begin{align}
&\lim_{t_1\to t_2}\left(1-e^{2i\omega(t_1-t_2)}\right)^{-1/2}\int \ud  \phi^{\q}\; e^{-A^{\q}(\phi^{\q})^2-B^{\q}\phi^{\q}}=\sqrt{\frac{\pi}{2}}\int{\rm d}\phi^{\q}\;\delta\left(\sqrt{\frac{\omega}{2\hbar}}\phi^{\q}\right)=\sqrt{\frac{\pi\hbar}{\omega}}~,
\end{align}
as above.

%%%%%%%%%%%%%%%%%%%%%%%%%%%%%%%%%%%%%%%%%%%%%%%%%%%%%

\section{Path integral for the dynamical part}
\label{sec:appendix_a}

For the dynamical part $e^{-\AI}$, we have the continuum expression given in Eq.~\eqref{eq:L}.
However, to gain a better understanding of the real-time path integral, it is convenient to consider the discrete form
\begin{align}
\AI=-\frac{i}{\hbar} \int \ud^{d} \mathbf{x}\sum_{j=1}^{m-1} \Bigg[
\phi^{\q}_j(\mathbf{x})\frac{\overline{\overline{\phi^{\cl}_{j+1}}}(\mathbf{x})-\phi^{\cl}_{j+1}(\mathbf{x})}{\ud t} 
-\ud t \sum_{n=1}^{+\infty} \frac{\left(\phi^{\q}_j(\mathbf{x})\right)^{2n+1}}{ 2^{2n}(2n+1)!} V^{(2n+1)}\Bigg]~,
\label{eq:dis}
\end{align}
where we have adopted the shorthand notation\footnote{The derivatives of the potential are understood as functions of $\phi^{\cl}_j(\mathbf{x})$. The expression here is tantamount to Eq.~(2.11) in Ref.~\cite{Mou:2019gyl}.
We point out, however, that there is a typo there: In Eqs.~(2.11) and~(2.15) of Ref.~\cite{Mou:2019gyl}, the sign in front of the $\lambda$ term should be positive.}%
\begin{align}
\overline{\overline{\phi^{\cl}_{j+1}}}(\mathbf{x})\equiv
2\phi^{\cl}_j(\mathbf{x})-\phi^{\cl}_{j-1}(\mathbf{x})-\ud t^2 \left(
-\bm{\nabla}^2\phi^{\cl}_j(\mathbf{x})
+V^{(1)}
\right)~.
\label{eq:overline}
\end{align}
We first notice that there are only linear terms in $\phi^{\cl}_{m}(\mathbf{x})$, where $t_m$ is the latest time on the contour. Thus, if we integrate all $\phi^{\cl}_{m}(\mathbf{x})$ out, we will obtain the functional delta functions $\delta\left(\phi^{\q}_{m-1}(\mathbf{x})\right)$. We can then proceed to integrate out all $\phi^{\q}_{m-1}(\mathbf{x})$, and the result will have an exponent similar to Eq.~\eqref{eq:dis}, but now with $t_{m-1}$ as the end point.
This corresponds to the contraction of the contour along the real-time line, and it is for this reason that the $\phi^{\q}$ will play no role in the spatial Bell inequalities in Sec.~\ref{sec:Bell}. For further details, see Refs.~\cite{Mou:2019tck, Mou:2019gyl}.

On the other hand, without the higher-order derivatives of the potential, e.g., in a free theory, we can integrate out all $\phi^{\q}$ simply via
\begin{align}\label{eq:app:fourier}
\int \ud \phi_j^{\q}(\mathbf{x})\; \exp\left(-\frac{i}{\hbar} \ud^d\mathbf{x}\, \phi_j^{\q}\, \frac{\phi_{j+1}^{\cl}(\mathbf{x})-\overline{\overline{\phi_{j+1}^{\cl}}}(\mathbf{x})}{\ud t}\right)
= \frac{2\pi \hbar \ud t}{\ud^d\mathbf{x}} ~\delta\left(\phi_{j+1}^{\cl}(\mathbf{x})-\overline{\overline{\phi_{j+1}^{\cl}}}(\mathbf{x})\right)~.
\end{align}
Given the shorthand notation \eqref{eq:overline}, the delta function above simply enforces the classical trajectory for $\phi^{\cl}_{j+1}$, given $\phi^{\cl}_{j}$ and $\phi^{\cl}_{j-1}$. We also note that we interpret Eq.~\eqref{eq:app:fourier} in the sense that space has also been discretized, with Eq.~\eqref{eq:app:fourier} referring to a particular spatial point $\mathbf{x}$. With this delta function, we can further integrate out the $\phi^{\cl}$ fields. This is how we proceed with the denominator in passing from Eq.~\eqref{eq:C_QFT_q_cl} to Eq.~\eqref{eq:QFTnew}.

For the numerator, we need to proceed a little differently, due to the presence of external operators. For sites on which none of the operators are located, we can still apply the integral above, which leads to the same delta functions. On the special sites at $t_1$ (denoted with a discrete index $j_1$), where an operator is situated, we apply the following integral:
{\small
\begin{align}\label{eq:app:fourier_sign}
&\int \ud \phi_{j_1}^{\q}(\mathbf{x}_1) \exp\left(\!-\frac{i}{\hbar}\,\ud^d\mathbf{x}\,\phi_{j_1}^{\q}(\mathbf{x}_1) \frac{\phi_{j_1+1}^{\cl}(\mathbf{x}_1)-\overline{\overline{\phi_{j_1+1}^{\cl}}}(\mathbf{x}_1)}{\ud t}\right)\nonumber\\&\qquad\times\frac{1}{2}\left[{\rm sign}\left(\phi^{\cl}_{j_1}(\mathbf{x}_1)+\frac{\phi^{\q}_{j_1}(\mathbf{x}_1)}{2}\right)+{\rm sign}\left(\phi^{\cl}_{j_1}(\mathbf{x}_1)-\frac{\phi^{\q}_{j_1}(\mathbf{x}_1)}{2}\right)\right]
\nonumber\\&\qquad\qquad=\frac{2\hbar \ud t}{\ud^d\mathbf{x}} \frac{
\sin\left(\frac{2\phi^{\cl}_{j_1}(\mathbf{x}_1)\ud^d\mathbf{x}}{\hbar \ud t}\left(\phi_{j_1+1}^{\cl}(\mathbf{x}_1)-\overline{\overline{\phi_{j_1+1}^{\cl}}}(\mathbf{x}_1)\right)\right)}
{\phi_{j_1+1}^{\cl}(\mathbf{x}_1)-\overline{\overline{\phi_{j_1+1}^{\cl}}}(\mathbf{x}_1)}~.
\end{align}}%
As with Eq.~\eqref{eq:app:fourier}, Eq.~\eqref{eq:app:fourier_sign} arises from discretizing space, with the equation holding for a particular position $\mathbf{x}_1$. In comparison to the denominator, a delta function
\begin{equation}
\delta\left(\phi_{j_1+1}^{\cl}(\mathbf{x}_1)-\overline{\overline{\phi_{j_1+1}^{\cl}}}(\mathbf{x}_1)\right)
\end{equation}
has been replaced by
\begin{equation}
\frac{1}{\pi}\frac{
\sin\left(\frac{2\phi^{\cl}_{j_1}(\mathbf{x}_1)\ud^d\mathbf{x}}{\hbar \ud t}\left(\phi_{j_1+1}^{\cl}(\mathbf{x}_1)-\overline{\overline{\phi_{j_1+1}^{\cl}}}(\mathbf{x}_1)\right)\right)}
{\phi_{j_1+1}^{\cl}(\mathbf{x}_1)-\overline{\overline{\phi_{j_1+1}^{\cl}}}(\mathbf{x}_1)}~.
\end{equation}
Notice that this reduces to the same delta function in the limit $2\phi^{\cl}_{j_1}(\mathbf{x}_1)\ud^d\mathbf{x}/(\hbar \ud t)\to 0$.

We are now in a position to integrate out the remaining $\phi^{\q}$, which fall into two categories: those between $t_0$ and $t_1$, and those between $t_1$ and $t_2$. Each integration results in a delta function of the form in Eq.~\eqref{eq:app:fourier}, and so the remaining $\phi^{\cl}$ integrals (except for $\phi_{j_1+1}^{\cl}(\mathbf{x}_1)$) pick out the classical trajectories, one starting at $t_0$, and the other starting at $t_1$. As a result, we can integrate out all $\phi^{\cl}$ in the numerator, except $\phi_{j_1+1}^{\cl}(\mathbf{x}_1)$. If we now return to Eq.~\eqref{eq:C_QFT_q_cl} and define the double angle brackets $\langle\!\langle\dots\rangle\!\rangle$ to be a shorthand notation for the path integrals \emph{without} the $\phi^\pm_0$ integrals, we find
\begin{align}
&\frac{1}{2}\big\langle\!\big\langle\big[{\rm sign}(\phi_{j_1}^+(\mathbf{x}_1))+{\rm sign}(\phi_{j_1}^-(\mathbf{x}_1))\big] {\rm sign}(\phi^{\cl}_{j_2}(\mathbf{x}_2))\big\rangle\!\big\rangle 
\nonumber\\
&\qquad=\frac{1}{\pi}\int \ud \phi^{\cl}_{j_1+1}(\mathbf{x}_1)\; \frac{
\sin\left(\frac{2\tilde\phi_{j_1}^{\cl}(\mathbf{x}_1)\ud^d\mathbf{x}}{\hbar \ud t}\left(\phi_{j_1+1}^{\cl}(\mathbf{x}_1)-\tilde\phi_{j_1+1}^{\cl}(\mathbf{x}_1)\right)\right)}
{\phi_{j_1+1}^{\cl}(\mathbf{x}_1)-\tilde\phi_{j_1+1}^{\cl}(\mathbf{x}_1)}\, {\rm sign}(\tilde{\tilde{\phi}}_{j_2}^{\cl}(\mathbf{x}_2))
\nonumber\\
&\qquad =\frac{1}{\pi}\int \ud \pi^{\cl}_{j_1}(\mathbf{x}_1)\; \frac{
\sin\left(\frac{2\tilde\phi_{j_1}^{\cl}(\mathbf{x}_1)\ud^d\mathbf{x}}{\hbar }\left(\pi_{j_1}^{\cl}(\mathbf{x}_1)-\tilde\pi_{j_1}^{\cl}(\mathbf{x}_1)\right)\right)}
{\pi_{j_1}^{\cl}(\mathbf{x}_1)-\tilde\pi_{j_1}^{\cl}(\mathbf{x}_1)} {\rm sign}(\tilde{\tilde{\phi}}_{j_2}^{\cl}(\mathbf{x}_2))~,
\end{align}
where we have introduced the momentum fields
\begin{align}
\pi^{\cl}_{j_1}(\mathbf{x}_1)\equiv \frac{\phi^{\cl}_{j_1+1}(\mathbf{x}_1)-\tilde \phi^{\cl}_{j_1}(\mathbf{x}_1)}{\ud t}
\qquad \text{and}\qquad
\tilde\pi^{\cl}_{j_1}(\mathbf{x}_1)\equiv \frac{\tilde\phi^{\cl}_{j_1+1}(\mathbf{x}_1)-\tilde \phi^{\cl}_{j_1}(\mathbf{x}_1)}{\ud t}~.
\end{align}
The tilde refers to the classical trajectories with the initial data at $t_0$. The $\tilde{\tilde{\phi}}^{\cl}_{j_2}(\mathbf{x}_2)$ also satisfies the classical equation, but with the initial data at $t_1$, i.e., including $\pi^{\cl}_{j_1}(\mathbf{x}_1)$. Thus, along with the initial Wigner function, we obtain the expression given in Eq.~\eqref{eq:QFTnew}.

%%%%%%%%%%%%%%%%%%%%%%%%%%%%%%%%%%%%%%%%%%%%%%%%%%%%%

\end{document}